\documentclass{IEEEtran}
\usepackage{cite}
\usepackage{amsmath,amssymb,amsfonts}
\usepackage{algorithmic}
\usepackage{textcomp}
\usepackage{graphicx}
\usepackage{xcolor,colortbl}
\usepackage{color,soul}

\usepackage[dvips]{epsfig}

\usepackage[caption=false,font=footnotesize]{subfig}

\usepackage{url,array,multirow,booktabs}
\usepackage{siunitx}
\usepackage{lipsum}
\usepackage[mathscr]{euscript}
\usepackage{comment}
\usepackage{xcolor,colortbl}
\usepackage{color,soul}
\usepackage[utf8]{inputenc}
\usepackage[english]{babel} 
\usepackage{mwe}

\newcommand{\dtoprule}{\specialrule{1pt}{0pt}{0.4pt}%
            \specialrule{0.3pt}{0pt}{\belowrulesep}%
            }
\newcommand{\dbottomrule}{\specialrule{0.3pt}{0pt}{0.4pt}%
            \specialrule{1pt}{0pt}{\belowrulesep}%
            }
            
            \newcommand{\STRUT}{\rule{0in}{2.3ex}}
\usepackage{diagbox} 
\usepackage{kotex} 
\usepackage[flushleft]{threeparttable} 
\usepackage{lscape} 
\usepackage{tabularx} 
\usepackage{tabu} 

\def\BibTeX{{\rm B\kern-.05em{\sc i\kern-.025em b}\kern-.08em
    T\kern-.1667em\lower.7ex\hbox{E}\kern-.125emX}}
\begin{document}
\title{Beam Squint in Ultra-wideband mmWave Systems: \\RF Lens Array vs. Phase-Shifter-Based Array}{
\author{Sang-Hyun Park, \IEEEmembership{Student Member, IEEE}, Byoungnam Kim, \IEEEmembership{{Member}, IEEE}, Dong Ku Kim, \IEEEmembership{Senior Member, IEEE}, Linglong Dai, \IEEEmembership{Fellow, IEEE}, Kai-Kit Wong, \IEEEmembership{Fellow, IEEE}, and Chan-Byoung Chae, \IEEEmembership{Fellow, IEEE}
\thanks{S.-H. Park, D. K. Kim, C.-B. Chae are with Yonsei University, Seoul, Korea (e-mail: \{williampark,dkkim,cbchae\}@yonsei.ac.kr).}
\thanks{B. Kim is with Sensor View, Gyeonggi-do, Korea (e-mail: klaus.kim@sensor-view.com).}
\thanks{L. Dai is with Tsinghua University, Beijing, China (e-mail: daill@tsinghua.edu.cn).}
\thanks{K.-K. Wong is with University College London, UK. (e-mail: kai-kit.wong@ucl.ac.uk).}
\thanks{Note to Reviewers: Full demo video is also available for reviewers. Please check your manuscriptcentral account. }}
\maketitle

\begin{abstract}
In this article, we discuss the potential of radio frequency (RF) lens for ultra-wideband millimeter-wave (mmWave) systems. In terms of the beam squint, we compare the proposed RF lens antenna with the phase shifter-based array for hybrid beamforming. To reduce the complexities for fully digital beamforming, researchers have come up with RF lens-based hybrid beamforming. The use of mmWave systems, however, causes an increase in bandwidth, which gives rise to the beam squint phenomenon. We first find the causative factors for beam squint in the dielectric RF lens antenna. Based on the beamforming gain at each frequency, we verify that, in a specific situation, RF lens can be free of the beam squint effect. We use 3D electromagnetic analysis software to numerically interpret the beam squint of each antenna type. Based on the results, we present the degraded spectral efficiency by system-level simulations with 3D indoor ray tracing. Finally, to verify our analysis, we fabricate an actual RF lens antenna and demonstrate the real performance using a mmWave, NI PXIe, software-defined radio system.
\end{abstract}

\begin{IEEEkeywords}
Beam squint, mmWave, lens antenna, hybrid beamforming, analog beamforming, software-defined radio, 3D ray tracing.
\end{IEEEkeywords}
\vspace{-5pt}
\section{Introduction}
\label{sec:introduction}
Fifth-generation (5G) cellular networks suggest a paradigm shift from the sub-6~GHz systems to millimeter wavelength (mmWave, 30–300~GHz). 
In the technical reports~\cite{TR19}, the 3rd~Generation Partnership Project (3GPP) has already announced the 3~GHz bandwidth~(26.5--29.5~GHz). An ultra-wideband range, the 3~GHz bandwidth has not been used in the sub-6~GHz frequency spectrum. Owing to the unique characteristics of mmWave, using it in cellular networks is different from using the sub-6~GHz bands. The mmWave band, for instance, is vulnerable to obstacles and large-distance path losses. This vulnerability has prompted researchers to use multiple-input multiple-output (MIMO) systems in the mmWave band to restore capacity and link reliabilities. 
Due to the shorter wavelengths of mmWave systems, more antenna elements can be integrated for the same array size. For fully digital beamforming (BF)~\cite{GJ17}, though, integrating more antenna elements raises, exponentially, the computational complexity.

Furthermore, as antenna elements require dedicated radio frequency (RF) chains~\cite{HGR+16}
, the hardware complexity increases as well. 
The application of phase-shifting to multiple data streams after the baseband to form a highly directed beam is called analog-digital hybrid BF. The hybrid BF achieves array and multiplexing gains simultaneously. 
In general, the array gain is proportional to the array size and the number of antenna elements. Between performance and hardware cost, though, the hybrid BF is subject to a trade-off. According to~\cite{BBS13} and~\cite{ZZ16}, the RF lens antennas can reduce the hardware complexity without incurring performance degradation. In the proposed lens-based hybrid BF system, the RF lens is placed at the front end of the antenna array (see~Fig.~\ref{FigureBQ}). Study results verify that the lens antenna, while reducing hardware complexity, achieves the conventional high gain and directivity with lens-focusing characteristics. 



\begin{figure*}[t]
\vspace{10pt}
  \centerline{\resizebox{1.9\columnwidth}{!}{\includegraphics{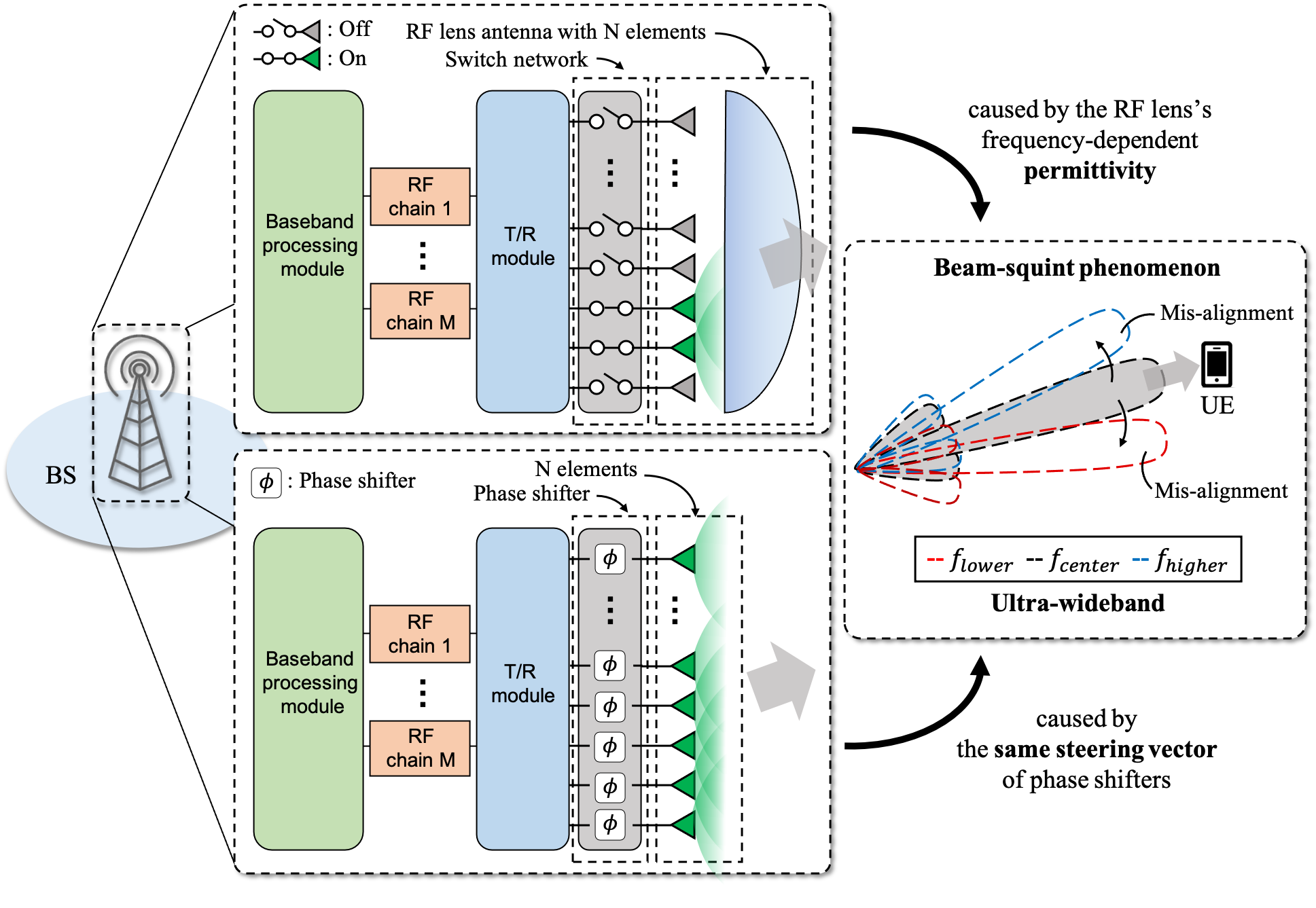}}}
   \caption{The hybrid BF system structures and the beam squint phenomenon of the dielectric RF lens and the phase shifter-based array. $f_\text{lower}$, $f_\text{center}$, and $f_\text{higher}$ denote lower frequency, center frequency, and higher frequency, respectively.}
   \label{FigureBQ}
\vspace{-10pt}
\end{figure*}
In a fully digital BF system, beam squint can be mitigated by frequency-dependent steering vectors via wideband beamforming structures such as tapped delay line, finite impulse response (FIR), and infinite impulse response (IIR) filters~\cite{LW10}. However, in hybrid analog–digital BF systems, the digital precoding process burden is divided into analog BF processing burden, and it is difficult to mitigate beam squint due to a narrowband characteristic of phase shifts~\cite{FBC+21}. Therefore a major problem remaining for hybrid BF, though, is beam squint. Beam squint refers to the phenomenon wherein the beam-steering angle is altered depending on the operating frequency~\cite{FBC+21}. The phase of each antenna element must be shifted to steer the analog BF in a specific direction. Because the wave number influences the array factor of the phased array, the operating frequency band alters the analog BF angle. 
Fig.~\ref{FigureBQ} shows the RF lens and phase shifter-based ultra-wideband hybrid BF structure and the beam squint phenomenon. Beams misaligned due to beam squint (colored in red and blue) will cause the transmitter to miss the user equipment (UE). Since these beams are incredibly narrow in a large array hybrid BF system, the beam squint phenomenon can be a critical issue that affects the received power. The dielectric RF lens and the phased array both generate beam squint, though their causes differ. In the former, the cause is the frequency-dependent permittivity of the RF lens and in the latter the cause is the same steering vector of the phased array in different frequency band. We detail this in Section~\ref{reason_beamsquint}. 

We summarize the contributions of this study as follows:
\begin{itemize}
\item We propose a low-cost practical solution to resolve the beam squint problem with the dielectric RF lens.
\item We verify that the beam squint problem of the dielectric RF lens system, even when not considering the true time delay line, can be negligible when using stable permittivity materials.
\item By analyzing the beam squint phenomenon in an ultra-wideband dielectric RF lens antenna via 3D electromagnetic simulation, we provide an insight into the causative factor and suppression method of beam squint in a lens-based beamforming system.
\item We analyze the relationship between each causative factor and beam squint by categorizing the causative factors as an inherent causative factor and an external factor.
\item These results are evaluated through an indoor 3D ray tracing system-level simulation and a link-level demonstration with a fabricated dielectric RF lens.
\end{itemize}


\vspace{-10pt}
The remainder of this article is organized as follows. Section~\ref{Sec.2} presents the inherent causative factors of beam squint in the RF lens and the phased array. Section~\ref{CST} presents the system model used in the 3D electromagnetic analysis software. In addition, we introduce the BF gain ratio that assesses the beam squint effects. We also classify and summarize the causative factors for the beam squint phenomenon. In Section~\ref{Sec.4}, we present the results of an indoor, map-based 3D ray-tracing method for accurate analysis of the beam squint. We then verify the degraded spectral efficiency and received~power~ratio of the fabricated RF lens through the NI mmWave software-defined radio (SDR) platform over the air. Finally, we summarize our results, contributions, and propose future works.

\begin{table*}[t]
\caption{Angle distortion values [$^{\circ}$] in Evaluation model~1 (EM1) and Evaluation model~2 (EM2)}
\label{table_20AD}
\centering
\begin{threeparttable}
{\begin{tabular}{c|cccccccccccc}
\dtoprule
\multicolumn{1}{c|}{\textbf{Phased array}} & \multicolumn{2}{c}{\textbf{27~GHz}} & \multicolumn{2}{c}{\textbf{27.5~GHz}} & \multicolumn{2}{c}{\textbf{28~GHz}} & \multicolumn{2}{c}{\textbf{29~GHz}} & \multicolumn{2}{c}{\textbf{29.5~GHz}} & \multicolumn{2}{c}{\textbf{30~GHz}} \\ \hline
\multicolumn{1}{c|}{{AoD$^{\dagger}$}}\STRUT & \multicolumn{1}{c}{{\textit{EM1.}}} & \multicolumn{1}{c|}{{\textit{EM2.}}} & \multicolumn{1}{c}{{\textit{EM1.}}} & \multicolumn{1}{c|}{{\textit{EM2.}}} & \multicolumn{1}{c}{{\textit{EM1.}}} & \multicolumn{1}{c|}{{\textit{EM2.}}} & \multicolumn{1}{c}{{\textit{EM1.}}} & \multicolumn{1}{c|}{\textit{{EM2.}}} & \multicolumn{1}{c}{\textit{{EM1.}}} & \multicolumn{1}{c|}{{\textit{EM2.}}} & \multicolumn{1}{c}{{\textit{EM1.}}} & {\textit{EM2.}} \\ 
6 deg\STRUT & -0.18 & \multicolumn{1}{c|}{} & -0.12 & \multicolumn{1}{c|}{} & -0.06 & \multicolumn{1}{c|}{} & 0.06 & \multicolumn{1}{c|}{} & 0.11 & \multicolumn{1}{c|}{} & 0.16 &  \\
12 deg & -0.55 & \multicolumn{1}{c|}{} & -0.36 & \multicolumn{1}{c|}{} & -0.18 & \multicolumn{1}{c|}{} & 0.17 & \multicolumn{1}{c|}{} & 0.33 & \multicolumn{1}{c|}{} & 0.48 &  \\
18 deg & -0.92 & \multicolumn{1}{c|}{*} & -0.60 & \multicolumn{1}{c|}{*} & -0.30 & \multicolumn{1}{c|}{*} & 0.28 & \multicolumn{1}{c|}{*} & 0.55 & \multicolumn{1}{c|}{*} & 0.82 & * \\
24 deg & -1.32 & \multicolumn{1}{c|}{} & -0.87 & \multicolumn{1}{c|}{} & -0.43 & \multicolumn{1}{c|}{} & 0.40 & \multicolumn{1}{c|}{} & 0.80 & \multicolumn{1}{c|}{} & 1.18 &  \\
30 deg & -1.76 & \multicolumn{1}{c|}{} & -1.15 & \multicolumn{1}{c|}{} & -0.57 & \multicolumn{1}{c|}{} & 0.54 & \multicolumn{1}{c|}{} & 1.06 & \multicolumn{1}{c|}{} & 1.56 &  \\
\midrule \midrule
\multicolumn{1}{c|}{\textbf{RF lens}} & \multicolumn{2}{c}{\textbf{27~GHz}} & \multicolumn{2}{c}{\textbf{27.5~GHz}} & \multicolumn{2}{c}{\textbf{28~GHz}} & \multicolumn{2}{c}{\textbf{29~GHz}} & \multicolumn{2}{c}{\textbf{29.5~GHz}} & \multicolumn{2}{c}{\textbf{30~GHz}} \\ \hline
\multicolumn{1}{c|}{{AoD$^{\dagger}$}}\STRUT & \multicolumn{1}{c}{{\textit{EM1.}}} & \multicolumn{1}{c|}{{\textit{EM2.}}} & \multicolumn{1}{c}{{\textit{EM1.}}} & \multicolumn{1}{c|}{{\textit{EM2.}}} & \multicolumn{1}{c}{{\textit{EM1.}}} & \multicolumn{1}{c|}{{\textit{EM2.}}} & \multicolumn{1}{c}{{\textit{EM1.}}} & \multicolumn{1}{c|}{\textit{{EM2.}}} & \multicolumn{1}{c}{\textit{{EM1.}}} & \multicolumn{1}{c|}{{\textit{EM2.}}} & \multicolumn{1}{c}{{\textit{EM1.}}} & {\textit{EM2.}} \\ 
6 deg \STRUT& -1.19 & \multicolumn{1}{c|}{} & -0.03 & \multicolumn{1}{c|}{} & -0.97 & \multicolumn{1}{c|}{} & -0.81 & \multicolumn{1}{c|}{} & -0.13 & \multicolumn{1}{c|}{} & -0.16 &  \\
12 deg & -0.37 & \multicolumn{1}{c|}{} & -0.11 & \multicolumn{1}{c|}{} & -0.48 & \multicolumn{1}{c|}{} & -0.46 & \multicolumn{1}{c|}{} & ~0.15 & \multicolumn{1}{c|}{} & -0.47 &  \\
18 deg & ~0.03 & \multicolumn{1}{c|}{**} & -0.01 & \multicolumn{1}{c|}{**} & -0.18 & \multicolumn{1}{c|}{**} & -0.30 & \multicolumn{1}{c|}{**} & -0.16 & \multicolumn{1}{c|}{**} & -0.20 & ** \\
24 deg& -0.23 & \multicolumn{1}{c|}{} & ~0.01 & \multicolumn{1}{c|}{} & -0.20 & \multicolumn{1}{c|}{} & -0.11 & \multicolumn{1}{c|}{} & -0.23 & \multicolumn{1}{c|}{} & -0.09 &  \\
30 deg& ~0.12 & \multicolumn{1}{c|}{} & ~0.61 & \multicolumn{1}{c|}{} & ~0.35 & \multicolumn{1}{c|}{} & ~0.50 & \multicolumn{1}{c|}{} & ~0.20 & \multicolumn{1}{c|}{} & ~0.49 & \\
\dbottomrule
\end{tabular}
}
\end{threeparttable}
\vspace{10pt}
\centering

\label{table_20PD}
\textsc{Power difference values [dBi] in Evaluation model~1 (EM1) and Evaluation model~2 (EM2)}\\
\vspace{5pt}
\begin{threeparttable}
\begin{tabular}{c|cccccccccccc}
\dtoprule
\multicolumn{1}{c|}{\textbf{Phased array}} & \multicolumn{2}{c}{\textbf{27~GHz}} & \multicolumn{2}{c}{\textbf{27.5~GHz}} & \multicolumn{2}{c}{\textbf{28~GHz}} & \multicolumn{2}{c}{\textbf{29~GHz}} & \multicolumn{2}{c}{\textbf{29.5~GHz}} & \multicolumn{2}{c}{\textbf{30~GHz}} \\ \hline
\multicolumn{1}{c|}{{AoD$^{\dagger}$}} \STRUT& \multicolumn{1}{c}{{\textit{EM1.}}} & \multicolumn{1}{c|}{{\textit{EM2.}}} & \multicolumn{1}{c}{{\textit{EM1.}}} & \multicolumn{1}{c|}{{\textit{EM2.}}} & \multicolumn{1}{c}{{\textit{EM1.}}} & \multicolumn{1}{c|}{{\textit{EM2.}}} & \multicolumn{1}{c}{{\textit{EM1.}}} & \multicolumn{1}{c|}{\textit{{EM2.}}} & \multicolumn{1}{c}{\textit{{EM1.}}} & \multicolumn{1}{c|}{{\textit{EM2.}}} & \multicolumn{1}{c}{{\textit{EM1.}}} & {\textit{EM2.}} \\ 
6 deg\STRUT &~0.4 & \multicolumn{1}{c|}{} &~0.2 & \multicolumn{1}{c|}{} & -0.0 & \multicolumn{1}{c|}{} &~0.0 & \multicolumn{1}{c|}{} &~0.2 & \multicolumn{1}{c|}{} & ~0.4 &  \\
12 deg &~0.5 & \multicolumn{1}{c|}{} &~0.2 & \multicolumn{1}{c|}{} & -0.1 & \multicolumn{1}{c|}{} &~0.1 & \multicolumn{1}{c|}{} &~0.2 & \multicolumn{1}{c|}{} &~0.4 &  \\
18 deg &~0.6 & \multicolumn{1}{c|}{0} &~0.2 & \multicolumn{1}{c|}{0} &~0.1 & \multicolumn{1}{c|}{0} &~0.0 & \multicolumn{1}{c|}{0} &~0.2 & \multicolumn{1}{c|}{0} &~0.3 & 0 \\
24 deg &~0.7 & \multicolumn{1}{c|}{} &~0.3 & \multicolumn{1}{c|}{} &~0.1 & \multicolumn{1}{c|}{} &~0.0 & \multicolumn{1}{c|}{} &~0.2 & \multicolumn{1}{c|}{} &~0.3 &  \\
30 deg &~1.0 & \multicolumn{1}{c|}{} &~0.5 & \multicolumn{1}{c|}{} &~0.2 & \multicolumn{1}{c|}{} & -0.0 & \multicolumn{1}{c|}{} & -0.0 & \multicolumn{1}{c|}{} &~0.2 &  \\
\midrule \midrule
\multicolumn{1}{c|}{\textbf{RF lens}} & \multicolumn{2}{c}{\textbf{27~GHz}} & \multicolumn{2}{c}{\textbf{27.5~GHz}} & \multicolumn{2}{c}{\textbf{28~GHz}} & \multicolumn{2}{c}{\textbf{29~GHz}} & \multicolumn{2}{c}{\textbf{29.5~GHz}} & \multicolumn{2}{c}{\textbf{30~GHz}} \\ \hline
\multicolumn{1}{c|}{{AoD$^{\dagger}$}}\STRUT & \multicolumn{1}{c}{{\textit{EM1.}}} & \multicolumn{1}{c|}{{\textit{EM2.}}} & \multicolumn{1}{c}{{\textit{EM1.}}} & \multicolumn{1}{c|}{{\textit{EM2.}}} & \multicolumn{1}{c}{{\textit{EM1.}}} & \multicolumn{1}{c|}{{\textit{EM2.}}} & \multicolumn{1}{c}{{\textit{EM1.}}} & \multicolumn{1}{c|}{\textit{{EM2.}}} & \multicolumn{1}{c}{\textit{{EM1.}}} & \multicolumn{1}{c|}{{\textit{EM2.}}} & \multicolumn{1}{c}{{\textit{EM1.}}} & {\textit{EM2.}} \\ 
6 deg \STRUT&~0.1 & \multicolumn{1}{c|}{} &-0.2 & \multicolumn{1}{c|}{} &~0.2 & \multicolumn{1}{c|}{} &~0.4 & \multicolumn{1}{c|}{} &~0.2 & \multicolumn{1}{c|}{} &~0.4 &  \\
12 deg &~0.0 & \multicolumn{1}{c|}{} &-0.1 & \multicolumn{1}{c|}{} & -0.4 & \multicolumn{1}{c|}{} & -0.3 & \multicolumn{1}{c|}{} & ~0.3 & \multicolumn{1}{c|}{} & -0.2 &  \\
18 deg&~0.3 & \multicolumn{1}{c|}{$^{\ddagger}$} &~0.0 & \multicolumn{1}{c|}{$^{\ddagger}$} &~0.0 & \multicolumn{1}{c|}{$^{\ddagger}$} &-0.4 & \multicolumn{1}{c|}{$^{\ddagger}$} &~0.1 & \multicolumn{1}{c|}{$^{\ddagger}$} &-0.3 & $^{\ddagger}$ \\
24 deg &~0.1 & \multicolumn{1}{c|}{} &-0.2 & \multicolumn{1}{c|}{} &-0.3 & \multicolumn{1}{c|}{} &-0.6 & \multicolumn{1}{c|}{} & -0.3 & \multicolumn{1}{c|}{} & -0.7 &  \\
30 deg &-0.6 & \multicolumn{1}{c|}{} &-0.2 & \multicolumn{1}{c|}{} &-0.9 & \multicolumn{1}{c|}{} & -0.1 & \multicolumn{1}{c|}{} & -0.1 & \multicolumn{1}{c|}{} &~0.9 & \\
\dbottomrule
\end{tabular}
{
        \begin{tablenotes}
\item[] Please refer to Section~\ref{Sec.3.1} and~\ref{Sec.3.2} for the definitions of angle distortion, power difference, EM1, and EM2.
        \item[$^{\dagger}$] indicates Tx beamforming direction in the azimuth angle of antenna via analog BF.
            \item[*] indicates that the values are not changed from \textit{Evaluation model~1} to \textit{Evaluation model~2}. 
            \item[**] $<$ .001
            \item[$^{\ddagger}$] $<$ .01
            \end{tablenotes}
        }
     \end{threeparttable}
     \vspace{-10pt}
\end{table*}

\section{Existing Suppression Methods and Beam squint in the RF Lens Antenna}
\label{Sec.2}
\subsection{Existing Beam squint Suppression Methods}
\label{Sec.222}
This section introduces the existing beam squint suppression method. In systems that use phased array structure, several methods have been proposed to solve the beam squint~\cite{FBC+21,WCW+18}. There are two primary methods: the method that uses true time delay (TTD) circuits and the method that does not use TTD circuits. The TTD method can eliminate the beam squint phenomenon by replacing the characteristic of the phase shifter with a TTD circuit, but the hardware cost for the implementation of TTD is not suitable for large-scale arrays~\cite{WCW+18}. In comparison, the method that does not use TTD cannot fundamentally eliminate beam squint impairments and only mitigates degradation performance caused by beam squint. The authors of \cite{FBC+21} proposed methods using beam broadening and TTD circuits in a hybrid beamforming system. In the first method, the authors proposed a beam broadening that divides the array into sub-arrays to widen the beamwidth. This mitigates the beam squint without structural changes, but the throughput performance deteriorates as the beam is broadened. In the second method, the beam squint is removed by adding TTD lines between the phase shifter and the RF chain. In particular, the second method shows a result close to fully digital BF in the ultra-wideband, but the cost of analog parts due to the addition of TTD still exists. In conclusion, these solutions are still subject to trade-offs between performance and hardware cost.

In the field of RF lens antenna, the authors in~\cite{GDZ+19} considered beamspace channel estimation for the mmWave wideband. By exploiting the effect of beam squint caused by the wideband, it was revealed that each path component of the wideband beamspace channel shows a unique frequency-dependent sparse structure. However, the causative factors of beam squint in the RF lens antenna and the resulting throughput degradation are not well understood.

As another solution, there is a method that use a lens that considers TTD properties. For example, a well-known ideal Rotman lens consists of transmission delay lines and eliminates beam squint with its lens shape, as it considers the TTD property between the beam port contour and antenna array. However, in practice, it is impossible to simultaneously design all beam ports in the Rotman lens strictly according to the geometric shape equation~\cite{LZ18}. The geometric shape equation makes the lens feasible for all signals with various and continuous directions. The TTD transmission delay lines of the practical Rotman lens are fabricated to support finite angular signals. Therefore, constrained imperfection of the Rotman lens results in remaining beam squint impairment. Motivated by this, we study the causes of the beam squint phenomenon and its performance assessment for an RF lens antenna made of dielectric material, not a specially designed lens for TTD.

\begin{figure*}[t]
\centering
{{\resizebox{1.30\columnwidth}{!}{\includegraphics{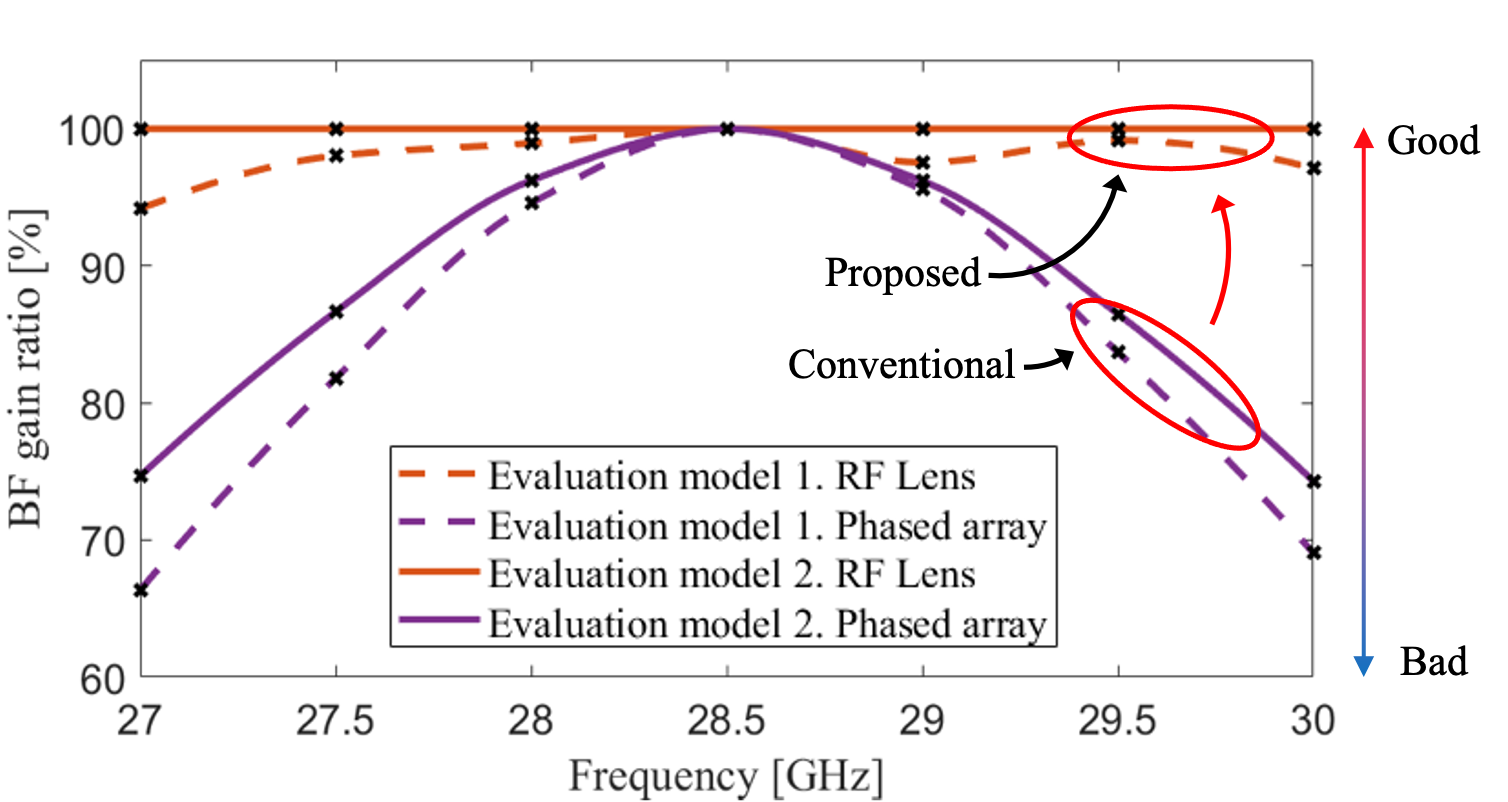}}}}
    \caption{The BF gain ratio along the frequency caused by the beam squint from the 3D electromagnetic simulation.} 
     \label{FigureDPBQ_Dia20}
     
\end{figure*}

\subsection{Why Does Beam squint Occur in the Lens Antenna Structure?}
\label{reason_beamsquint}
Beam squint is generated by the material permittivity of the dielectric RF lens, which is a function of operating frequency. In an ultra-wideband system, the RF lens antenna demonstrates beam squint for reasons different from those for the phased array.
Based on Snell’s law, an electromagnetic wave is refracted when passing through a dielectric lens. According to the Drude–Lorentz model~\cite{Lorentz}, the refraction is caused by the fact that permittivity is a function of frequency -- one of the causative factors of beam squint. Permittivity is also not linear owing to the damping constant and resonance frequency. Permittivity consists of real and imaginary parts. The real part is the dielectric constant, and the tangent loss is the ratio of the real part to imaginary part.

What accounts for the beam refraction, as it can be observed, is the real part of the refractive index. Because the permittivity influences the real part of the refractive index, the wave with other frequency bands passes through the material with different degrees of refraction, resulting in beam squint at the lens. The permittivity affects the refractive index, so it is advantageous to have fewer dielectric-constant-and-tangent-loss variations in the operating frequency band. Therefore, it is essential to select the lens material with fewer permittivity variations in the lens-based BF system.

\subsection{Lens Antenna Structure}
\vspace{-1pt}
As shown in Fig.~\ref{FigureBQ}, the RF lens antenna consists of two parts: the antenna array connected with RF chains and the lens structure in the front end of the antenna array. The lens structure is used to focus the beam generated by the antenna array for an analog BF. Beam steering is achieved using the lens refractions with the on-off switching of the antenna~\cite{CSK+18, KLM+16}. To reduce hardware complexity and allow faster beam tracking that is faster than phase-shifted array, it is possible to use switching BF using one element or a subset of the antenna array~\cite{ZZ16, CSK+18}. 

According to Section~\ref{reason_beamsquint}, the refractive index of the lens is a function of frequency. When selecting the material to manufacture the lens, one should take care to account for the beam squint phenomenon. In designing the RF lens, our previous study considered five materials--Teflon type A, polyethylene, polycarbonate, Boron Nitride, and MgO. We need to consider whether these materials have stable dielectric constants and tangent loss values. However, for most of the RF lens materials in the mmWave band, precise dielectric constant and tangent loss measurements data are not available. In this article, the lens material used in the simulation is Teflon type~A. According to~\cite{O15}, Teflon type A has small loss tangent values and no large deviations in the permittivity from the 27~GHz to the 30~GHz range. These characteristics of Teflon work in favor of the lens to mitigate beam squint. In this article, we use a hyperbolic curve shape of the lens based on~\cite{KLM+16}.

\begin{table*}
\caption{A category of causative factors for beam squint in Evaluation model~1 and 2.}
\label{category}
\centering
\begin{threeparttable}
\begin{tabular}{cc|c|c|cc}
\hline \hline
\multicolumn{2}{c|}{\multirow{2}{*}{\textbf{Category of causative factors\STRUT}}} & \multirow{2}{*}{\textbf{Causative factor \STRUT}} & \multirow{2}{*}{\textbf{Affected parameters\STRUT}} & \multicolumn{2}{c}{\multirow{2}{*}{\textbf{\begin{tabular}[c]{@{}c@{}}Evaluation model\STRUT\end{tabular}}}} \\
\multicolumn{2}{c|}{} &  &  & \multicolumn{2}{c}{} \\ \midrule
\multicolumn{1}{c|}{\multirow{5}{*}{\begin{tabular}[c]{@{}c@{}}Inherent factors$^{*}$\STRUT\\ of the \textbf{ABF} structures\end{tabular}}} & \multirow{2}{*}{\begin{tabular}[c]{@{}c@{}}Inherent factors\STRUT\\ of the \textbf{phased array} structure\end{tabular}} & Array size\STRUT & HPBW & \multicolumn{1}{c|}{\multirow{7}{*}{\textbf{\begin{tabular}[c]{@{}c@{}}Evaluation\\ model 1\end{tabular}}}} & \multirow{5}{*}{\textbf{\begin{tabular}[c]{@{}c@{}}Evaluation\\ model 2\end{tabular}}} \\ \cline{3-4}
\multicolumn{1}{c|}{} &  & Steering vector\STRUT & Angle distorsion & \multicolumn{1}{c|}{} &  \\ \cline{2-4}
\multicolumn{1}{c|}{} & \multirow{3}{*}{\begin{tabular}[c]{@{}c@{}}Inherent factors\STRUT\\ of the \textbf{RF lens} structure\end{tabular}} & Array size\STRUT & HPBW & \multicolumn{1}{c|}{} &  \\ \cline{3-4}
\multicolumn{1}{c|}{} &  & \multirow{2}{*}{\begin{tabular}[c]{@{}c@{}}Frequency-dependent\STRUT\\ permittivity\end{tabular}} & Angle distortion\STRUT & \multicolumn{1}{c|}{} &  \\ \cline{4-4}
\multicolumn{1}{c|}{} &  &  & Power difference\STRUT & \multicolumn{1}{c|}{} &  \\ \cline{1-4} \cline{6-6} 
\multicolumn{2}{c|}{\multirow{2}{*}{\begin{tabular}[c]{@{}c@{}}External factor\STRUT \\unrelated to the types of \textbf{ABF} structures\end{tabular}}} & \multirow{2}{*}{\begin{tabular}[c]{@{}c@{}}Narrowband antennas\STRUT\\ in the ULA$^{\dagger}$\end{tabular}} & Angle distortion\STRUT & \multicolumn{1}{c|}{} & \multirow{2}{*}{--} \\ \cline{4-4}
\multicolumn{2}{c|}{} &  & Power difference\STRUT & \multicolumn{1}{c|}{} &\\ \hline \hline 
\end{tabular}
{
        \begin{tablenotes}
 \item[$^{*}$] Inherent factors are related to the types of analog beamforming (ABF) structures.
            \item[$^{\dagger}$] The antenna gain of a narrowband antenna element in the ULA is not consistent for the ultra-wideband.
           
            \end{tablenotes}
        }
     \end{threeparttable}

\end{table*}

\section{Beam squint Analysis based on 3D Electromagnetic Analysis Software}
\label{CST}
In this section, to analyze the beam squint phenomenon in the RF lens and the phased array, we use the 3D electromagnetic simulation, CST Studio Suite. We introduce the assessment model in Sections~\ref{Sec.3.1} and \ref{Sec.3.2} and present our simulation results in Section~\ref{Sec.3.3}.

\vspace{-7pt}
\subsection{Beam squint Assessment Model}
\label{Sec.3.1}
For an accurate analysis of the beam squint, we set two evaluation models. Evaluation model~1 takes into account only \textit{Assumption~1}. Evaluation model~2 considers \textit{Assumptions~1}~and~{\textit{2}}. The assumptions are explained as follows: 

\textit{Assumption~1}: In an RF lens, the factor that affects beam refraction is only the refractive index, which is a function of permittivity. \textit{Assumption~2}: The antenna gain of the antenna elements in the uniform linear array (ULA) is consistent for the ultra-wideband. {\textit{Assumption~1} is made to eliminate the likelihood of intervention by other variables (e.g., diffraction phenomenon on the edge of RF lens) in the refractive index. {\textit{Assumption~2} removes the distorted components according to the narrowband antenna elements. In Evaluation model~2, we can observe the beam squint phenomenon due to the inherent causative factors, excluding external causative factors such as the narrowband antenna elements. Here, the inherent causative factor refers to the influence related to analog BF structures. The external causative factor refers to the influence unrelated to analog BF structures, such as the influence from the antenna element.
\vspace{-5pt}
\subsection{Parameters and Degraded Power Caused by Beam squint for Analysis}
\label{Sec.3.2}
As beam squint occurs, the BF gain is degraded. This degraded gain leads to a degraded received power. We analyze, from the perspective of the transmitter, the extent of the attenuated BF gains. To assess the beam squint phenomenon, we introduce several parameters. 
The first is angle distortion~(AD). AD is the difference between the azimuth angle at peak BF gain for the operating frequency and the center frequency. The second parameter is power difference (PD). PD represents the difference between the peak BF gain at each operating frequency and the center frequency. The third parameter is the half-power beamwidth (HPBW) at each frequency band. The HPBW depends on the array size, with a larger array size resulting in a narrower beam, leading to critical mis-alignment. Evaluating beam squint by AD alone cannot accurately account for the performance attenuation. Hence, we add PD and HPBW to provide the BF gain ratio and received power ratio that evaluate more general and accurate beam squint performance.


Based on the direction of the analog beam at the center frequency, the BF~gain~ratio is defined as the percentage of the ratio of the directivity gain at the center frequency to the mis-aligned directivity gain at the operating frequency. For example, an 80$\%$ BF~gain~ratio value at 27~GHz implies that the antenna gain at 27~GHz in the direction aligned with 28.5~GHz is 80$\%$ of the antenna gain at 28.5~GHz.

\vspace{-5pt}
\subsection{AD, PD, and BF~Gain~Ratio Comparison for Phased Array with RF Lens in a ULA}
\label{Sec.3.3}
We simulated two different antenna structures--an RF lens and a phased array in 20$\lambda$ antenna diameters, where $\lambda$ is a wavelength at the center frequency. The 20$\lambda$ arrays consist of 28x1 patch antennas with $\frac{\lambda}{2}$ distance between two adjacent patch antennas. Table~\ref{table_20AD}, based on a 3D electromagnetic analysis software simulation, shows the AD and PD data of the RF lens antennas and the phased array. In Table~\ref{table_20AD}, the y-axis represents the angle of departure (AoD) of Tx analog BF, and the x-axis represents operating frequencies from 27~GHz to 30~GHz without 28.5~GHz. The data in the tables consist of AD and PD values introduced in Section~\ref{Sec.3.2}. We analyze Table~\ref{table_20AD} data as the phased array, RF lens antenna, and BF~gain~ratio. 
Finally, with Table~\ref{category}, we present a summary of the categories of the causative factors and affected parameters in each antenna structure. \\

\vspace{-10pt}
\textbf{Phased array:}
According to Evaluation model~1 data, shown above in Table~\ref{table_20AD}, AD in the phased array becomes more distorted as the frequency gets further away from the center frequency. This is because the steering vector set for the center frequency causes angle distortion at different operating frequencies. In addition, AD becomes more distorted as the AoD increases. Hence, maximum distortion values are -1.76$^\circ$ (30$^\circ$, 27~GHz) and 1.56$^\circ$ (30$^\circ$, 30~GHz). It is worth noting that the data from Evaluation model~2 are the same as those from Evaluation model~1. In addition, the existing PD values in Evaluation model~1 become zero in Evaluation model~2. These data imply that PD in a phased array is affected by a narrowband antenna beam pattern, the only external factor. In the phased array, however, AD does not depend on that external factor.\\

\vspace{-10pt}
\textbf{RF lens antenna:}
According to Evaluation model~1, we can see that the AD values of the RF lens are not consistent. For those causative factors of distortion, we can see that the data from Evaluation model~2--where AD values of RF lens are less than 0.001--are both independent of AoD and frequencies. This data analysis implies that most of the causative factors which cause AD are inconsistent antenna gains across the frequency band. In Table~\ref{table_20AD} (below), the PD of the RF lens has the same result. The PD values of Evaluation model~1 in the RF lens are less than 0.01. This result implies that the most most significant causative factor of PD is not an RF lens's inherent factor but rather the narrowband antennas as an external factor. The remaining values are caused by the frequency-dependent permittivity mentioned in Section~\ref{reason_beamsquint}.\\

\begin{figure*}[t!]
\centering
{{\resizebox{2\columnwidth}{!}{\includegraphics{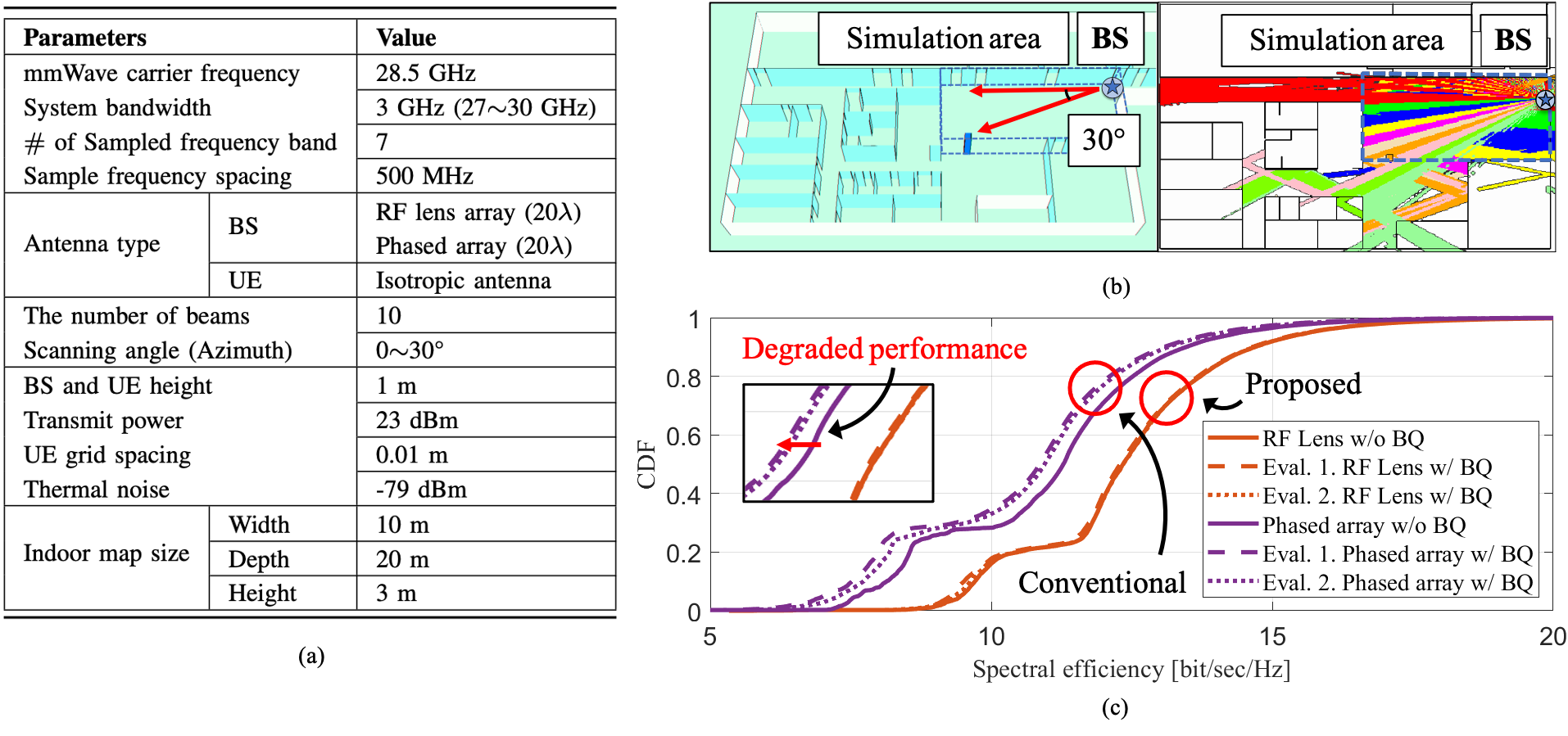}}}}
\vspace{-5pt}
\caption{(a) Parameters in the system-level simulation; (b) The indoor map of 3D ray tracing system-level simulation. Each color means the best beams which are selected using the Best Beam Selection method among the ten beams (0 $\sim$ 30$^{\circ}$); (c)~Degraded spectral efficiency assessment due to beam squint for the conventional and proposed antennas.}
     \label{SLS_total}
     \end{figure*}

     \begin{figure*}[t]
\centering
{{\resizebox{2\columnwidth}{!}{\includegraphics{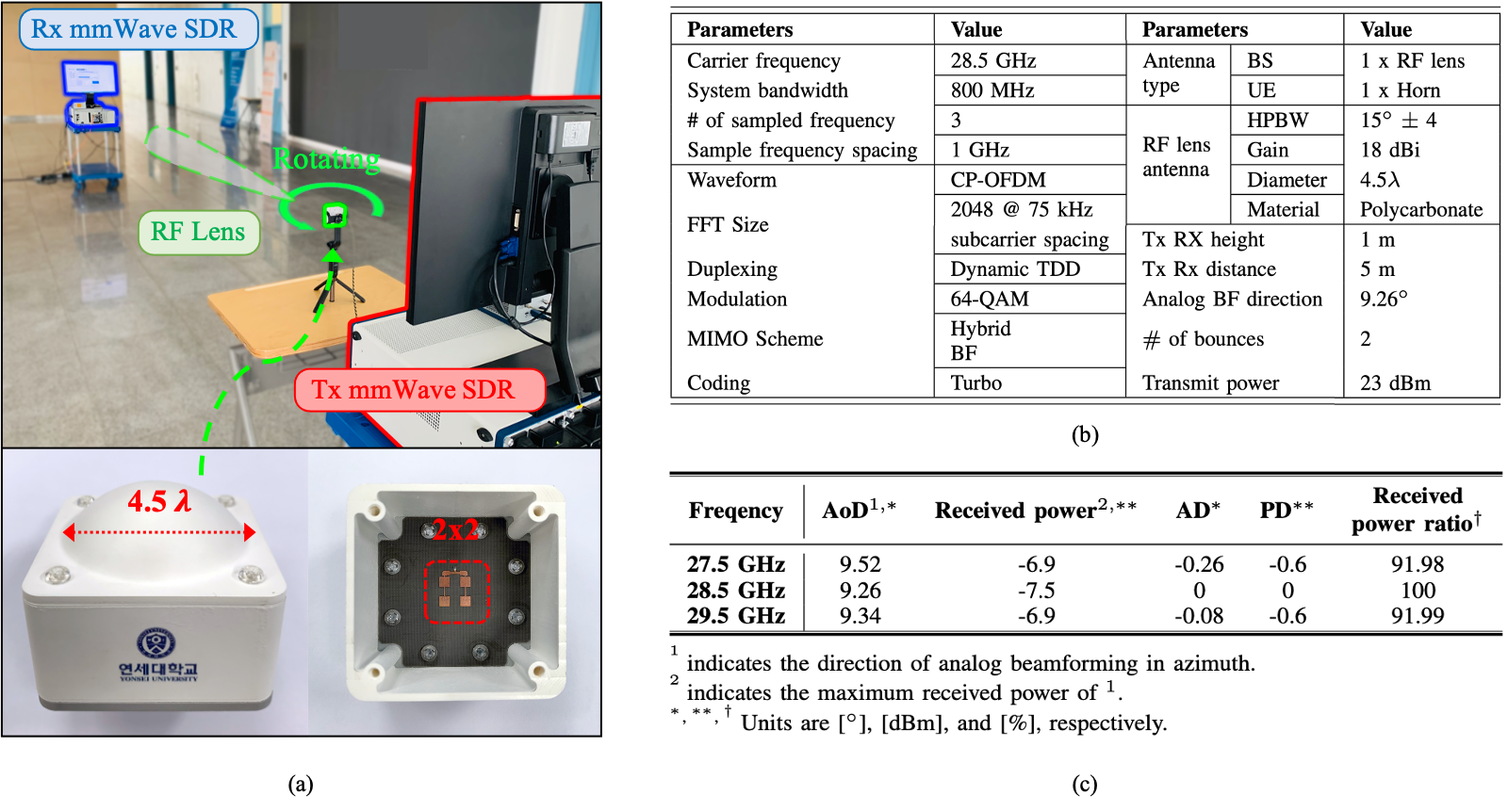}}}}
    \caption{(a) Indoor mmWave software-defined radio (SDR) testbed with a fabricated RF lens for beam squint; (b) Parameters in the link-level demonstration; (c) Beam squint results of the fabricated RF lens via mmWave hardware testbed}
     \label{Link_demo_total}
     \end{figure*}	 
     
\vspace{-10pt}
\textbf{A summary of the causative factors in each antenna: }
In Table~\ref{table_20AD}, we analyze and summarize, based on the data, the causative factors and the parameters (AD, PD and HPBW). In Table~\ref{category}, the causative factors are classified as either inherent or external. The inherent factor concerns a structural feature of the antenna that gives rise to the beam squint phenomenon. Evaluation model~1 contains two types of causative factors, and Evaluation model~2 analyzes only the inherent factors. An inherent factor of the phased array structure is the steering vector that affects AD. In the RF lens structure, a causative factor is the frequency-dependent permittivity, and it affects AD and PD. In the external factor category, a narrowband antenna affects the PD of the phased array and the AD and PD of the RF lens. The external factor constitutes--especially in the RF lens--the most significant causative factor. Hence, the RF lens array that uses ideal or ultra-wideband antenna elements is a robust answer to the beam squint problem.\\

\vspace{-10pt}
\textbf{BF~gain~ratio:}
We analyze degraded BF gain by converting the Table~\ref{table_20AD} data to BF gain ratio. Fig.~\ref{FigureDPBQ_Dia20} presents the BF gain ratio values of each Evaluation method in terms of the operating frequency. The solid lines signify Evaluation model~1; the dashed lines signify Evaluation model~2. The BF gain ratio from the proposed RF lens is much less degraded than that of the conventional phased array. In Evaluation model~1, we can see that the phased array has only 67$\%$ and 69$\%$ BF gain of the standard BF gain (at the center frequency) at 27~GHz and 30~GHz, respectively. In Evaluation model~1, by contrast, a minimum value of the RF lens’s BF gain ratios has 93$\%$. By excluding any external factor in Evaluation model~2, the RF lens is almost free of BF gain attenuation by the beam squint.\\

\vspace{-15pt}
\section{System-Level and Link-Level Assessments}
\label{Sec.4}
This section presents the evaluations of the spectral efficiency and received~power~ratio at the system and link-levels. Note that the system-level simulation is based on the antenna patterns of CST simulation in Section~\ref{CST}. Furthermore, we use the received power ratio instead of the BF gain ratio. The received power ratio is defined as the ratio of received powers based on distorted AD, PD, and HPBW, similar to the BF gain ratio.
\subsection{System-Level Performance Analysis and Evaluation}
\label{Sec_SLS}
We evaluate the RF lens-embedded system and phased array with a system-level simulator using Wireless System Engineering (WiSE)--a 3D ray tracing tool developed by Bell Labs~\cite{KLM+16}. 
For realistic performance evaluations, we model a building in a university, as shown in Fig.~\ref{SLS_total}(b). The simulation area is indoors with a line-of-sight (LoS) environment to consider the beam squint phenomenon and straightness characteristics of the mmWave signal. With a multiuser environment, we assume perfect channel state information (CSI) in all Rx locations. We adopt the best beam selection (BBS) method to choose the best beam for each user according to received power. As shown in Table~\ref{table_20AD}, because the AD unit is 0.01$^{\circ}$, we set a 0.01~m distance for accurate received power and high resolution of AoD estimations. 
In Fig.~\ref{SLS_total}(b), the BS (marked by a blue star) steers ten beams between 0$^{\circ}$ and 30$^{\circ}$ towards the left in the simulated space. To estimate the received~power~ratio, and spectral efficiency, researchers aggregate the received power for the other frequencies of a selected beam at each location. In the two Evaluation models, we analyze the two types of antennas. The detailed system parameters are given in Fig.~\ref{SLS_total}(a).

In the simulations, each user selects the best beam based on the center frequency at each UE position. However, though the beam was selected based on the center frequency, the overall received power is attenuated by ADs and PDs at the other operating frequencies, depending on the beam squint phenomenon. Fig.~\ref{SLS_total}(c) presents the spectral efficiencies for the system-level simulation. In the spectral efficiency graphs, the solid line indicates the spectral efficiency without the beam squint phenomenon. Evaluation model~1 is indicated by dashed lines; Evaluation model~2 by dotted lines.
In Fig.~\ref{SLS_total}(c), the phased array in Evaluation model~1 (represented as the dashed line) demonstrates a 1.11 dB degradation relative to the scenario without beam squint (represented as the solid line). Furthermore, the phased array in Evaluation model~2 shows a degradation of 0.87~dB. In the case of the RF lens, the deterioration in system performance is minimal. Evaluation model~1 shows a degradation of 0.12~dB, while Evaluation model~2 has no deterioration. It is worth noting that the spectral efficiency of the lens antenna system is higher at 2 bit/s/Hz than that of the phased array. Although the same power is used in both systems, the lens antenna with the beam-focusing characteristic performs twice as good as the phased array.

In conclusion, it is impossible in the hybrid BF system to entirely compensate for the AD and PD of the phased array without increasing hardware complexity. The phased array structure, therefore, has an inherent limitation due to the beam squint phenomenon. In contrast, the beam squint problem of the RF lens system can, with stable permittivity materials, be negligible.

\subsection{Link-Level Performance Analysis and Evaluation}
\subsubsection{System Configuration}
To test the lens antenna performance as a component of a transmission system, we evaluated its link-level performance using the mmWave transceiver system software-defined radio (SDR) platform~\cite{SHP+20}. The communication system consists of a LabVIEW system design software and PXIe SDR.
As shown in Fig.~\ref{Link_demo_total}(a), the Tx node (as a BS) and Rx node (as a UE) are separated by 5m. The Tx consists of an IF-LO module, I/Q-generator, and Tx NI up-converter.
The Rx consists of an IF-LO module, I/Q-digitizer, decoder, multiple-access channel, and Rx downconverter.
The fabricated lens antenna is used in the BS with uplink and downlink frequencies of 28.5~GHz and 800 MHz bandwidth each. We test the link-level performance when the lens antenna is a transmitter, and the feed horn antenna is a receiver. A single data stream is transmitted and received via a modulation and coding scheme, more specifically, a 64-quadrature amplitude modulation.

The detailed parameters are given in Fig.~\ref{Link_demo_total}(b). Fig.~\ref{Link_demo_total}(a) shows the fabricated RF lens antenna used in the BS. The fabricated lens material is polycarbonate, not Teflon type A. The diameter of the RF lens is 4.5$\lambda$. Because we use a different RF lens, the results in Fig.~\ref{Link_demo_total}(c) are different from the system-level simulation results in Section~\ref{Sec_SLS}. However, this demonstration is performed in the same environment as that used for the system-level simulation. We verify the beam squint phenomenon of the fabricated lens in terms of its AD, PD, and received~power~ratio. In this demonstration, the analog BF’s AoD is 9.26$^{\circ}$. We measure the received powers at 27.5~GHz, 28.5~GHz, and 29.5~GHz.
     
\subsubsection{MmWave Link-Level Beam squint Performance Assessment}
Here, the received~power~ratio values are empirically demonstrated for assessing performance degradation due to beam squint in the indoor link-level testbed. The testbed verifies the existence of beam squint in the fabricated lens antenna. As a result, Fig.~\ref{Link_demo_total}(c) illustrates the AD, PD, and received~power~ratio from a beam squint perspective. The AoD changes are 0.26$^{\circ}$ at 27.5~GHz and 0.08$^{\circ}$ at 29.5~GHz. The HPBW of the fabricated lens is 15$^{\circ}$. Regarding the received power ratio, this RF lens shows a percentage of 91.98$\%$ and 91.99$\%$ at 27.5~GHz and 29.5~GHz, respectively.

\section{{\fontsize{11}{14}\selectfont CONCLUSION}}
\label{Sec.Conclusion}
This study investigated the beam squint phenomenon of a phased array and an RF lens antenna in an ultra-wideband mmWave system. Our work analyzed, in a dielectric RF lens-based hybrid BF system, the causative factors for the beam squint problem, separated into inherent causative factors and an external factor. Using a 3D electromagnetic analysis software, we compared and analyzed the phased array and the lens antenna under the same conditions in terms of beam squint with the BF gain and the received power. Through the analysis, we verified that the beam squint problem of the RF lens system can, with stable permittivity materials, be made negligible. Then, using 3D ray tracing in an indoor environment, we showed the degraded spectral efficiency, caused by beam squint, for both the phased array and the RF lens antenna. Finally, we demonstrated the indoor mmWave link-level with the fabricated RF lens and verified the performance degradation incurred by beam squint. In future work, we will evaluate frequency-dependent permittivity variances in terms of beam squint extent in an ultra-wideband BF system.


\bibliographystyle{ieeetr}
\bibliography{CommuMag_squint}

\end{document}